\documentclass[twocolumn,preprintnumbers, amssymb,amsmath,aps,floatfix,prd,nofootinbib,superscriptaddress,showpacs]{revtex4-1}

\usepackage{epsfig}
\usepackage{bm}
\usepackage{amssymb}
\usepackage{amsmath}
\usepackage{color}
\usepackage{subfigure}
\usepackage[colorlinks,
            linkcolor=blue,
            anchorcolor=black,
            citecolor=blue
            ]{hyperref}

\newcommand{\beq}{\begin{eqnarray}}
\newcommand{\eeq}{\end{eqnarray}}

\begin{document}
\title{Sub-threshold $J/\psi$ and $\Upsilon$ Production in $\gamma A$ Collisions}

\author{Yoshitaka Hatta}
\affiliation{Physics Department, Building 510A, Brookhaven National Laboratory, Upton, NY 11973}

\author{Mark Strikman}
\affiliation{Pennsylvania State University, University Park, PA, 16802}
\affiliation{Physics Department, Building 510A, Brookhaven National Laboratory, Upton, NY 11973}

\author{Ji Xu}
\affiliation{Department of Physics and Astronomy, Shanghai Jiao Tong University, Shanghai, China}
\affiliation{Nuclear Science Division, Lawrence Berkeley National
Laboratory, Berkeley, CA 94720, USA}

\author{Feng Yuan}
\affiliation{Nuclear Science Division, Lawrence Berkeley National
Laboratory, Berkeley, CA 94720, USA}

\begin{abstract}
We study  sub-threshold heavy quarkonium ($J/\psi$ and $\Upsilon$) photo-productions in $\gamma A$ collisions as an independent test of the universality of the nucleon-nucleon short range correlation (SRC) in nuclear scattering processes. Just below the $\gamma p$ threshold, the cross section is dominated by the mean field contribution of nucleons inside the nucleus. The SRC contributions start to dominate at lower photon energies, depending on the fraction of the SRC pairs in the target nucleus. We give an estimate of the cross sections in the sub-threshold region both for $J/\psi$ and $\Upsilon$. This may be helpful for future measurements at JLab as well as at the Electron-Ion Collider in  the U.S., and especially in China.
\end{abstract}
\maketitle

\section{Introduction}

Nucleon-nucleon short range correlation (SRC) describes an important aspect of the nuclear structure and has been a subject of intensive studies in the last few decades, see, for example, Refs.~\cite{Frankfurt:2008zv,Arrington:2011xs,Hen:2013oha,Hen:2016kwk,Fomin:2017ydn,Cloet:2019mql}. In the configuration space, the SRC represents the pair of nucleons (predominantly proton-neutron) that are close to each other, whereas in momentum space, they have large relative momentum but small total momentum. Recent experimental efforts from JLab have stimulated much research interest, by making connections between the SRC and the EMC effect in nuclear structure function measurements~\cite{Egiyan:2005hs,Seely:2009gt,Weinstein:2010rt,Fomin:2011ng,Hen:2012fm,Arrington:2012ax,Hen:2014nza,Cohen:2018gzh,Duer:2018sby,Duer:2018sxh,Schmookler:2019nvf}.  

The universality of the SRC is an important underlying feature in all these studies. Universality is the statement that the SRC is responsible for the EMC effect across different nuclei in the same manner. In particular, it has been demonstrated that the nuclear structure functions of different nuclei in the EMC region become a universal function once they are appropriately rescaled by the number of SRC pairs~\cite{Schmookler:2019nvf,Segarra:2019gbp}. To fully establish the physics case of the connection, we need to build a rigorous test for this universality, especially, from gluonic processes. This is because the existing EMC effects in these studies mainly focus on the quark sector of the partonic structure of the nucleus. It is of crucial importance to demonstrate the existence and the universality of the SRC contributions in the gluonic sector as well.  

In a recent paper by two of us~\cite{Xu:2019wso}, it was suggested that heavy flavor production in deep inelastic scattering (DIS) can provide a gluonic probe for the SRC contribution and further test the above mentioned universality. This can be achieved by measuring the charm-structure function in $eA$ collisions, and at the same time, one can also measure the so-called sub-threshold heavy flavor production in $\gamma A$ collisions. A crude model assumption has been applied to estimate the sub-threshold $J/\psi$ production in $\gamma A$ collisions. In this paper, we will perform a detailed calculation. In particular, we will include not only the SRC contributions, but also the mean field contributions. The latter contributions are important for heavy quarkonium production near the threshold. By comparing these two contributions, we will be able to pinpoint the energy range where the SRC contributions are dominant and can be applied to study the universality of SRC contributions. 

Sub-threshold hadron production in nuclear scattering processes has a long history, see, for example, the strangeness production in $pA$ and $AA$ collisions~\cite{Randrup:1980qd,Aichelin:1986ss,Schnetzer:1982ga,Schnetzer:1989vy,Adamczewski-Musch:2018xwg}, anti-proton in $pA$ collisions~\cite{Carroll:1988fc,Shor:1990qw}. In a recent experimental effort~\cite{Bosted:2008mn}, the sub-threshold $J/\psi$ production in $\gamma A$ collisions has been investigated at JLab with a low incoming photon energy (around $5.7\rm GeV$ in the rest frame of the nuclear target). No observation was reported. This is consistent with our estimate because the cross section is too small, see the detailed discussions in Sec.~III.

A number of experiments have already been approved to carry out $J/\psi$ production in various photo-nuclear experiments.  We expect that these experiments will measure $J/\psi$ production in the sub-threshold region, and provide an important test of the SRC universality. Meanwhile, there has been a strong proposal to build an intermediate energy electron-ion collider in China (EicC)~\cite{NuXu}, where the energy range is ideal to study both near and sub-threshold $\Upsilon$ production in $\gamma p$ and $\gamma A$ collisions, respectively. There is also a possibility to study the near-threshold production of $\Upsilon$ at the EIC in the U.S. \cite{Lomnitz:2018juf} and at RHIC using ultraperipheral collisions \cite{Hatta:2019lxo}. Our calculations in this paper will provide an important guidelines for these future experiments.

The rest of our paper is organized as follows. In Section~II, we take an example of $J/\psi$ production in photon-deuteron collisions to illustrate the method of our calculations. In Section~III, we extend that into  generic $eA$ collisions. Section~IV is devoted to $\Upsilon$ production. Based on these results, in Section~V we propose a simple scaling relation as a test of the universality of the SRC. We finally  summarize our paper in Section~VI.

\section{Sub-threshold $J/\psi$ production in $\gamma d$ Collisions}

Near-threshold photo-production of $J/\psi$  \cite{Gittelman:1975ix,Camerini:1975cy,Ali:2019lzf}  has attracted a lot of attention lately due to its connection to the proton mass problem \cite{Kharzeev:1998bz,Brodsky:2000zc,Frankfurt:2002ka,Gryniuk:2016mpk,Hatta:2018ina,Hatta:2019lxo,Mamo:2019mka}, see also, \cite{Miller:2015tjf}. For a nucleon target, the threshold photon energy for the reaction $\gamma p\to J/\psi+p'$   is $E_\gamma\approx 8.2$ GeV in the nucleon rest frame. If the target is  a heavier nucleus, $J/\psi$ can be produced at lower energies. The rate of this reaction is sensitive to the mass of the target as well as the momentum distribution of nucleons inside the target. In this section, we consider a deutron ($d$) target and compute the cross section in the sub-threshold region $E_\gamma<8.2$ GeV. 

For a nucleon target, following \cite{Brodsky:2000zc,Xu:2019wso} we parameterize the total cross section by the so-called energy fraction parameter 
\begin{equation}
    \chi_{J/\psi}=\frac{M_{J/\psi}^2}{2E_\gamma M_p}+\frac{M_{J/\psi}}{E_\gamma}=\frac{M_{J/\psi}^2+2M_pM_{J/\psi}}{W_{\gamma p}^2-M_p^2} \ ,
\end{equation}
where $M_p$ and $M_{J/\psi}$ are the  nucleon and $J/\psi$ masses, respectively, and $W_{\gamma p}$ is the center-of-mass energy of the photon-nucleon system. The threshold limit   corresponds to $\chi_{J/\psi}\to 1$. In Ref.~\cite{Xu:2019wso}, it was found that the following simple parametrization gives a very good description of the latest experimental data from the GlueX collaboration at JLab~\cite{Ali:2019lzf}
\begin{equation}
    \sigma_{\gamma p\to J/\psi}(W_{\gamma p})= \sigma_0^{\gamma p}
    (1-\chi_{J/\psi})^\beta \ , 
\end{equation}
where the parameters  $\sigma_0=11.3~\rm nb$ and $\beta=1.3$ have been fitted to the data. This form is also consistent, at least near the threshold region, with the calculation based on the AdS/CFT correspondence~\cite{Hatta:2018ina,Hatta:2019lxo,Mamo:2019mka}. In Ref.~\cite{Xu:2019wso}, the authors further estimated the cross section in the deuteron target case by a simple substitution $\chi_{J/\psi}\to \tilde \chi_{J/\psi}=\frac{M_{J/\psi}^2}{2E_\gamma 2M_p}+\frac{M_{J/\psi}}{E_\gamma}$ \cite{Xu:2019wso}. This, of course, is a very crude estimate. In the following, we apply the impulse approximation to evaluate the sub-threshold production cross section through a convolution method for the deuteron. We then extend to other nuclei by applying the SRC universality. 

\begin{figure}[h]
\centering
\includegraphics[width=0.8\columnwidth]{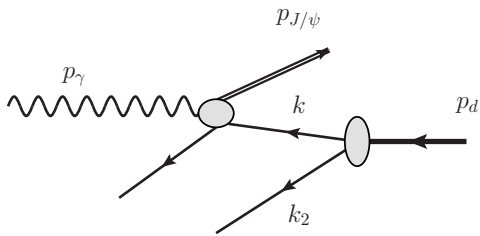}
\centering
\caption{Schematic diagram for $J/\psi$ production in $\gamma+d$ process.}
\label{deuterium}
\end{figure}

The reaction of interest is illustrated in Fig.~\ref{deuterium}. An incoming photon with momentum $p_\gamma=(E_\gamma,0_\perp,E_\gamma)$ interacts with one of the nucleons from the deuteron at rest and produces a  $J/\psi$ in the final state. The spectator carries momentum $k_2=\left(\sqrt{M_p^2+\vec{k}^2},-\vec{k}\right)$, and the center-of-mass energy squared of the photon-nucleon system that produces $J/\psi$ is 
\begin{equation}
    W_{\gamma p'}^2=(p_\gamma+p_d-k_2)^2 \ , \label{gwp}
\end{equation}
where  $p_d =(2M_p,\vec{0})$ is the  deuteron momentum. The struck nucleon has momentum $k=p_d-k_2=(\epsilon,\vec{k})$ with 
\beq
\epsilon=2M_p-\sqrt{M_p^2+\vec{k}^2}.
\label{as}
\eeq
In the impulse approximation we can write the total $J/\psi$ photo-production cross section off a deuteron target as
\begin{equation}
    \sigma_{\gamma d}=2\int d^3 k \rho_n(k) {\widetilde{\cal F}}(k)\sigma_{\gamma p}(W_{\gamma p'}) \ , \label{k}
\end{equation}
where a factor of 2 is included to take into account both the proton and neutron contributions. 
 $\rho_n(k)$  is the single-nucleon momentum distribution inside the deuteron.   This is normalized to unity $\int d^3k \rho_n(k)=1$, and describes  the probability distribution of the proton (or the neutron) carrying momentum $\vec{k}$ in the deuteron rest frame.  The distribution is dominated by the $S$-wave contribution at small $k$, whereas in the larger-$k$ region $k>k_F\sim 300$ MeV ($k_F$ is the Fermi momentum) it is dominated by the $D$-wave contribution.  The latter represents the SRC.  
$\tilde {\cal F}(k)$ accounts for the flux factor difference for $\sigma_{\gamma p}(W_{\gamma p'})$ from the free nucleon target case
\begin{equation}
    \widetilde{\cal F}(k)=\frac{E_\gamma(\epsilon-k^z) }{E_\gamma M_p}=\frac{2M_p-\sqrt{\vec{k}^2+M_p^2}-k^z}{M_p} \ . \label{from}
\end{equation} 
This arises because the incident nucleon is moving and not on-mass-shell.

For a given value of sub-threshold photon energy $E_\gamma<8.2$ GeV, the integration region in (\ref{k}) is determined by the condition $W_{\gamma p'}^2 >(M_p+M_{J/\psi})^2$. In principle, one also has to impose $\epsilon-k^z>0$ (see (\ref{from})), but this is automatically satisfied if the first condition is met. (Note that the important region is $k^z<0$.)
\begin{figure}[h]
\centering
\includegraphics[width=1.0\columnwidth]{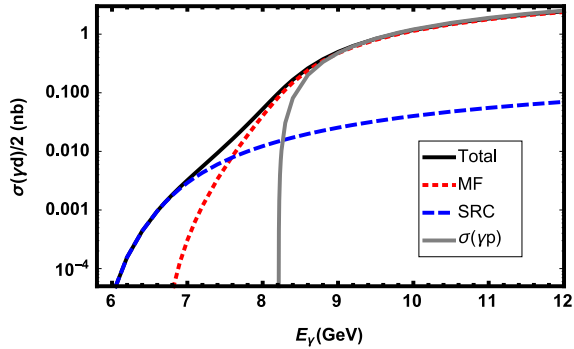}
\centering
\caption{Cross sections per nucleon for near and sub-threshold $J/\psi$ production in $\gamma d$ collisions as function of  $E_\gamma$ in the target rest frame. The near threshold cross section for $\gamma p$ collisions is also shown as a reference.}
\label{deuterium-sub}
\end{figure}
In Fig.~\ref{deuterium-sub}, we plot the near threshold and sub-threshold $J/\psi$ production cross section in $\gamma d$ collision calculated from the momentum distribution  in \cite{Wiringa:2013ala}. We also plot the near-threshold production in $\gamma p$ collisions as a reference. The mean field (MF) curve is the contribution from the low-momentum region defined here with a sharp momentum cutoff $k<300$ MeV. This improves the previous prediction in \cite{Xu:2019wso} and the energy dependence is now a smooth function around the threshold $E_\gamma\lesssim 8.2\rm$ GeV. However, the MF contribution drops dramatically around $E_\gamma\sim 7.5$ GeV below which  the SRC contribution ($k>300$ MeV) starts to dominate. This is easy to understand intuitively: Due to high intrinsic momentum $\vec{k}$ from the SRC, $J/\psi$ can be produced with smaller photon energies.  Around $E_\gamma\sim 7$ GeV, the cross section per nucleon is about $3.2$ pb, and is completely dominated by the SRC contribution. We thus conclude that, for the deuteron target, the interesting region to focus in future experiments is $E_\gamma\lesssim 7$ GeV. Events observed in this region provide an unambiguous signal for the gluonic probe of the SRC. We hope this will be carried out soon at JLab.

\section{SRC Universality in $\gamma A$ Collisions}

We now turn to the sub-threshold production of $J/\psi$ in $\gamma A$ collisions. A crucial difference from the deuteron case is that one now needs to introduce the so-called spectral function $\rho_A(k,\epsilon)$ and write   
\begin{equation}
    \sigma_{\gamma A}=A\int d^3 k d\epsilon\rho_A(k,\epsilon) {\widetilde{\cal F}}(k,\epsilon)\sigma_{\gamma p}(W_{\gamma p'}) \ , \label{int}
\end{equation}
where $\widetilde{\cal F}=(\epsilon-k^z)/M_p$ and 
\begin{eqnarray}
W_{\gamma p'}^2&=&(E_\gamma+\epsilon)^2-(E_\gamma+k^z)^2-\vec{k}_\perp^2  \nonumber\\
    &=&2E_\gamma \epsilon-2E_\gamma k^z+\epsilon^2-\vec{k}^2\ .
\end{eqnarray}
The point is that $\epsilon$ and $\vec{k}$ are now independent variables. 
Due to the collective effects and the shell-structure of the nucleus, the struck nucleon energy $\epsilon$ is not fixed as in (\ref{as}), but follows a smooth distribution.  This is accounted for by introducing the nuclear spectral function.  There is vast literature on the determination of the spectral functions from theory and experimental data, and one may use them to carry out the integral (\ref{int}). Here, however, we are particularly interested in the impact of the SRC. We therefore divide the cross section into the MF and SRC parts 
\begin{eqnarray}
\bar\sigma_{\gamma A}=\frac{\sigma_{\gamma A}}{A}&\equiv &\int_{k<k_{F}} \bar{\sigma}_{MF}(k)+\int_{k>k_F}\bar{\sigma}_{SRC}(k)\ ,
\end{eqnarray}
where $\bar\sigma$ stands for the cross section per nucleon, $k_F= 300$ MeV and make a separate approximation for $\rho_A$ in each region.

\subsection{Mean Field Contribution}

Let us start with the low-momentum, mean field contribution. In this case, the spectator is a nucleus with mass number $A-1$ and momentum $-\vec{k}$. Therefore, the scattering nucleon has energy,
\begin{eqnarray}
    \epsilon&=&M_A-\sqrt{(M_A-M_p+\Delta \epsilon)^2+\vec{k}^2}  \nonumber \\ 
    &\approx& M_p  -\frac{\vec{k}^2}{2M_p(A-1)}- \Delta \epsilon \ ,
\end{eqnarray}
where $M_A$ is the nucleus mass and $\Delta \epsilon$ is the separation energy of the  nucleon initially bound in shell model orbits. In principle, $\Delta \epsilon$ has a smooth distribution, but here we approximate it as a constant of order $10\sim 20$ MeV. Compared to this, the ${\cal O}(\vec{k}^2)$ term can be neglected for $A\gg 1$. We thus adopt a model  
\beq
  \rho_A^{(MF)}(k,\epsilon)  = \rho_A(k) \delta(\epsilon-M_p +\Delta \epsilon).
  \eeq
  We then parameterize the momentum distribution in the nucleus $\rho_A(k)$ in the Gaussian form
\begin{equation}
    \rho_A^{(MF)}(k)=\frac{P_{MF}}{N_0(k_F\sqrt{\pi} /2)^3}e^{-4k^2/k_F^2}\ ,\label{mfnormalization}
\end{equation}
where $P_{MF}$ represents the fraction of the mean field contribution to the momentum distribution. For a large nucleus, $P_{MF}$ is known to be about $80\%$. $N_0\approx 0.954$ is a factor needed to renormalize the distribution such that $P_{MF}=\int_{k<k_F} d^3 k\rho_A(k)$. Therefore, the mean field contribution can be written as
\begin{eqnarray}
    \bar\sigma_{MF}=\int_{k<k_F} d^3k\rho_A^{(MF)}(k){\widetilde{\cal F}}(k,\epsilon)\sigma_{\gamma p}(W_{\gamma p'}) \ ,
\end{eqnarray}
where $\epsilon=M_p-\Delta\epsilon$ with $\Delta\epsilon=0.02\rm GeV$ in the following numeric calculations. Again, we have applied the isospin symmetry and the above equation contains both proton and neutron contributions. 

\subsection{SRC Contribution from the Spectral Function}

The SRC contribution can be evaluated in a similar manner, but in this case the spectral function is much more involved than that for the MF contribution. It can be modeled by considering a pair of nucleons with high back to back momenta moving in the mean field~\cite{CiofidegliAtti:1991mm,CiofidegliAtti:1995qe,Weiss:2018tbu}. For example, the proton spectral function from the SRC in a nucleus can be written as \cite{Weiss:2018tbu}
\begin{eqnarray}
    \rho_A^{(SRC)p}(k,\epsilon)&=&C_{pn}^1S_{pn}^1(k,\epsilon)+C_{pn}^0S_{pn}^0(k,\epsilon)\nonumber\\
    &&+2C_{pp}^0S_{pp}^0(k,\epsilon) \ ,
\end{eqnarray}
where $C_{pn}^i$ and $C_{pp}^i$ represent the so-called nuclear contacts. They measure the probability to find $(pn)$ and $(pp)$ pairs with spin $i$ in the nucleus. The spectral functions are different for pairs with different quantum numbers. Moreover, the neutron spectral functions are different from the proton ones. Using isospin symmetry  $C_{nn}^0=C_{pp}^0$, we can write the total (proton and neutron) contribution from the SRC as 
\begin{eqnarray}
  &&  \bar{\sigma}_{SRC}=\int_{k>K_F} d^3k d\epsilon \widetilde{\cal F}(k,\epsilon)\sigma_{\gamma p}(W_{\gamma p'})  \left[C_{pn}^1S_{pn}^1(k,\epsilon)\right.\nonumber\\
    &&\quad +\left.C_{pn}^0S_{pn}^0(k,\epsilon)+C_{pp}^0\left(S_{pp}^0(k,\epsilon)+S_{nn}^0(k,\epsilon)\right)\right].
\end{eqnarray}

\begin{figure}[h]
\centering
\includegraphics[width=1.0\columnwidth]{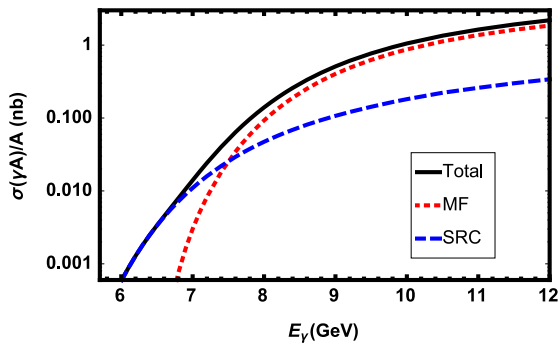}
\centering
\caption{Sub-threshold $J/\psi$ production in photon-Carbon collisions as a  function of incoming photon energy $E_\gamma$. We have estimated $P_{MF}=0.84$ for the mean field normalization in Eq.~(\ref{mfnormalization}).}
\label{nuclear-sub-spectral}
\end{figure}

We take the example of Carbon-12 and use the spectral functions~\cite{Duer:2018sxh,Jackson} 
calculated from the AV-18 potential with the following contacts:
\begin{eqnarray}
    C_{pp}^0 &=& C_{nn}^0 = 1.140 \%\ ,\nonumber\\
C_{pn}^0 &=& 1.244 \% \ ,\nonumber\\
C_{pn}^1 &=& 15.876 \% \ .   
\end{eqnarray}
In the numeric calculations, we normalize the spectral function in the full kinematics. Applying the isospin symmetry, we find that this normalization implies the following relation between the MF spectral function and the SRC spectral function, 
\begin{eqnarray}
    &&\int_{k<k_F} d^3 k \rho_A^{(MF)}(k)+\int_{k>K_F} d^3k d\epsilon  \left[C_{pn}^1S_{pn}^1(k,\epsilon)\right.\nonumber\\
    && \quad +\left.C_{pn}^0S_{pn}^0(k,\epsilon)+C_{pp}^0\left(S_{pp}^0(k,\epsilon)+S_{nn}^0(k,\epsilon)\right)\right] \nonumber\\
    &&=1.
\end{eqnarray}
From this normalization condition, we obtain the coefficient $P_{MF}=0.84$ for the MF fraction. The final result for the cross section per nucleon is shown in Fig.~\ref{nuclear-sub-spectral}. We can clearly see that again the SRC contribution dominates over the MF contribution in the kinematic region  $E_\gamma<7\,\rm GeV$. 

If we extend our calculation to $E_\gamma=5.7$ GeV where the previous JLab experiment~\cite{Bosted:2008mn}  searched for sub-threshold $J/\psi$ production, the cross section is about $0.05\, \rm pb$. The smallness of this value is likely the reason why no events were observed in this experiment.

\section{Sub-threshold $\Upsilon$ Production in $\gamma A$ Collisions}

The calculations in the last two sections can be straightforwardly extended to $\Upsilon$ production. However, currently experimental data for $\Upsilon$ production in $\gamma p$ collisions are not available in the threshold region $E_\gamma \approx 57$ GeV, or $\sqrt{s}=10.4$ GeV. (They are available only in the high energy region.)  Still, we can  give a rough estimate of the $\gamma p$ cross section by implementing necessary modifications to the formula for $J/\psi$.  

First of all, we argue that the functional form of energy dependence should be the same, because this only concerns the ``gluon" content in the nucleon.\footnote{We note, however, that very recently $\sigma^\Upsilon$ has been calculated in a holographic model \cite{Mamo:2019mka}. The result suggests that the threshold region may be very narrow in $E_\gamma$ and is quickly taken over by the asymptotic high energy (Pomeron) behavior.} Therefore, we can use the same parameterization for the cross section,
\begin{equation}
    \sigma_{\gamma p}^\Upsilon(W_{\gamma p})=\sigma_0^\Upsilon (1-\chi_\Upsilon)^{\beta_b}  \ ,
\end{equation}
where $\chi_\Upsilon$ is now defined as
\begin{equation}
    \chi_\Upsilon=\frac{M_\Upsilon^2+2M_\Upsilon M_p}{W_{\gamma p}^2-M_p^2} \ .
\end{equation}
 A major difference in the normalization $\sigma_0$ comes from the wave functions of  $J/\psi$ and $\Upsilon$ at the origin. This can be estimated from their respective leptonic decay widths, or as the ratio of photo-production cross sections at high energy evaluated at the same value of $x$. From the HERA experiments and recent measurements of photo-production of $J/\psi$ and $\Upsilon$ at the LHC, we find that the ratio between these two is about  $200$. 
 
In order to estimate the exponent $\beta_b$, we notice that part of the power behavior  $(1-\chi)^{\beta_b}$ comes from the phase space integral. Let us  assume that the differential cross section with respect to the momentum transfer $t$ has the following power behavior,
\begin{equation}
    \frac{d\sigma}{dt}\propto \frac{1}{(-t+\Lambda^2)^4} \ , \quad \sigma = \int_{t_{min}}^{t_{max}} \frac{d\sigma}{dt},
\end{equation}
where we choose $\Lambda\sim 1\,\rm GeV$. Because $t_{min}$ and $t_{max}$ depend on the center-of-mass energy and the quarkonium mass, this can generate different $\chi$-dependence in $\sigma$. A numerical estimate gives a relative factor of $0.6(1-\chi)^{0.5}$ between the $\Upsilon$ and $J/\psi$ cases.  
 We thus arrive at the following estimate
\begin{equation}
    \frac{\sigma_0^\Upsilon}{\sigma_0^{J/\psi}}\approx  
  \frac{0.6}{200}, ~~\beta_b=\beta+0.5=1.8 
  \ .
\end{equation}

\begin{figure}[h]
\centering
\includegraphics[width=1.0\columnwidth]{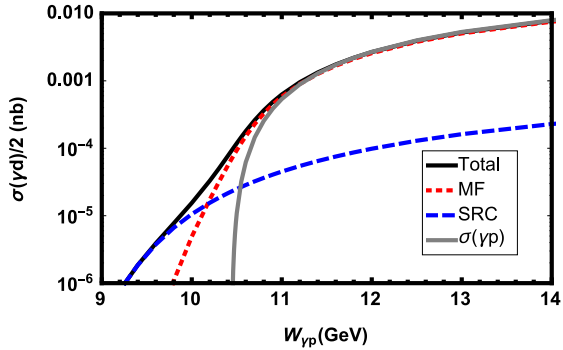}
\centering
\caption{The total cross section per nucleon for near and sub-threshold $\Upsilon$ production in  $\gamma d$ collisions as a function of the center of mass energy of photon-nucleon collisions $W_{\gamma p}$. The near threshold cross section for $\gamma p$ collisions is also shown as a reference.}
\label{upsilond}
\end{figure}

With these parameters, we compute the cross sections for $\Upsilon$ production in $\gamma p$ and $\gamma d$ collisions, and plot the result as a function of the center-of-mass energy of photon-nucleon collision $W_{\gamma p}$ in Fig.~\ref{upsilond}. This energy range is ideally suited for the EicC, but it could also be studied at the EIC in the U.S. \cite{Lomnitz:2018juf}, or  even at RHIC by focusing on the ultraperipheral $AA$ collisions (UPCs) \cite{Hatta:2019lxo}. We find that the $\Upsilon$ cross section near threshold is  about $10^{-3}\sim 10^{-2}$ nb, and this is further reduced to  $10^{-5}\, \rm nb$ around the region  $W_{\gamma p}\lesssim 9.7\, \rm GeV$  where the SRC contribution starts to dominate. It may be challenging to measure the cross section in this region.

\section{Discussions}

In light of our results in the previous sections, we update the prediction in  \cite{Xu:2019wso} as 
\begin{eqnarray}
    &&\left. \frac{\bar \sigma_{\gamma A\to J/\psi}}{ \bar\sigma_{\gamma d\to J/\psi}}\right|_{E_\gamma\sim 7\rm GeV}=\left.\frac{ \bar\sigma_{\gamma A\to \Upsilon}}{ \bar\sigma_{\gamma d\to \Upsilon}}\right|_{W_{\gamma p}\sim 9.7\rm GeV}\nonumber\\
    && =\frac{n_{src}^A/A}{n_{src}^d/2}=\left.\frac{F_2^A(x_B, Q^2)/A}{F_2^d(x_B, Q^2)/2}\right|_{1.4<x_B<1.8},
    \label{ratio}
\end{eqnarray}
where $n^A_{src}$ is the number of SRC pairs in nucleus $A$. The above ratio is also referred as $a_2^A=\frac{n_{src}^A/A}{n_{src}^d/2}$. The ratio of the structure functions $F_2$ has been measured in previous DIS experiments with nuclear targets. Future measurements of the sub-threshold cross section will be a clean test of the universality of the SRC in these nuclei. 

\begin{figure}[h]
\centering
\includegraphics[width=1.0\columnwidth]{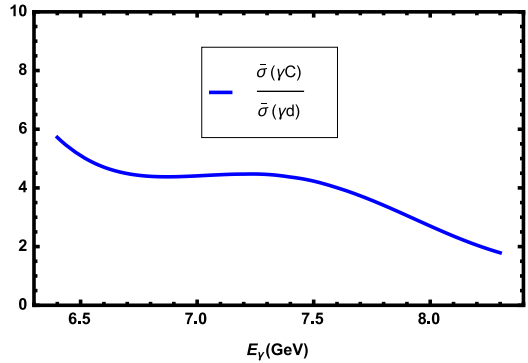}
\centering
\caption{Ratio of the $J/\psi$ photo-production cross sections per nucleon between the Carbon-12 target and deuteron target as a function of incoming photon energy $E_\gamma$ in the rest frame of the nuclear target. The plateau in this ratio indicates the onset of the SRC universality.}
\label{SRC-ratio}
\end{figure}

As an example, in Fig.~\ref{SRC-ratio}, we plot the cross section ratio between the Carbon-12 and deuteron targets as a function of incoming photon energy $E_\gamma$. From this plot, we can clearly see that the ratio increases from 2 around the $\gamma p$-threshold ($E_\gamma\sim 8.2~\rm GeV$) to a plateau behavior of $4.2$ around $E_\gamma\sim 7.4~\rm GeV$. Notice that the plateau starts already in the region $E_\gamma \sim 7.4\,{\rm GeV}$ where the mean field contribution is still significant. It blows up when $E_\gamma$ reaches the deuteron threshold around $5.6~\rm GeV$. Therefore, we conclude that the energy window to observe the SRC universality will be around $(6.5-7.4)~\rm GeV$.

Actually, this window turns out to be somewhat narrower than what one would expect from Eq.~(\ref{ratio}). It is known that the structure function ratio on the right hand side, when plotted as a function the light-cone momentum fraction of the interacting nucleon $\alpha=\alpha(x_B,Q^2)$ (defined in Eq.~(18) of Ref.~\cite{Frankfurt:1993sp}), exhibits a plateau for $1.3 < \alpha < 1.7$. One can consider a similar scaling variable $\alpha \sim 1/E_\gamma$ for the present problem, and this suggests that the plateau in $E_\gamma$ may actually be wider. Indeed, the present calculation may have large model uncertainties in the small $E_\gamma$ region where the denominator $\sigma_{\gamma d}$ becomes extremely small. In view of this, it is very interesting to see how low in $E_{\gamma}$ the plateau persists in future experimental data.

In our considerations we have neglected final state interactions of the produced $J/\psi$ and $\Upsilon$.
The analysis of $J/\psi-N$ absorption at somewhat higher energies suggests that effective  $J/\psi-N$ interactions in the discussed energy range  is of the order of a few mb, and hence in the first approximation they can be neglected for all but heaviest nuclei. Another possible effect is the dependence of the $J/\psi$ production cross section on the virtuality of the interacting nucleon. In principle, such an effect can be significant since the deviations from the many nucleon approximation are enhanced in SRC, see e.g.,
Refs.~\cite{Frankfurt:1988nt,Schmookler:2019nvf}. However this effect is expected to be the same  for SRC in different nuclei. So it should not modify the scaling relation
(\ref{ratio}).

\section{Conclusion}

In this paper, we have provided a detailed derivation of the sub-threshold heavy quarkonium production cross section in $\gamma A$ collisions.  We find that the sub-threshold cross section close to the threshold is dominated by the mean field effects, while in the deeply sub-threshold  region the dominant contribution comes from the SRC. In the latter region, the universality of the SRC can be tested. 

For  $J/\psi$ production relevant to the JLab kinematics, we found that the SRC contribution is dominant around the incoming photon energy  $E_\gamma =7.5\, \rm GeV$ and below in the rest frame of the nuclear target. The predicted cross sections are sizable and should be easily measured in the upcoming experiments. 
For $\Upsilon$ production, the required kinematical range is ideal for the EicC, but it could also be studied at the U.S. EIC and RHIC. However, the sub-threshold cross section is not as large as that for $J/\psi$, and this imposes a challenge in future measurements.

{\it Acknowledgement.} We thank Jackson Pybus for sending us the SRC spectral functions for Carbon-12. M.~S. would like to thank Brookhaven National Laboratory (BNL) for hospitality during the time this project started. His visit was supported by an LDRD from Brookhaven Science Associates (BSA). The work of Y.~H., J.~X. and F.~Y. is partially supported by the LDRD programs of BNL and Lawrence Berkeley National Laboratory, the U.S. Department of Energy, Office of Science, Office of Nuclear Physics, under contract numbers   DE-SC0012704, DE-AC02-05CH11231 and  DE-FG02-93ER40771.

\end{document}